\newif\ifArxiv
\Arxivtrue
\ifArxiv
\documentclass[10pt]{article}
\pdfoutput=1 

\usepackage{algorithm}
\usepackage{bigfoot}  
\usepackage{graphicx}
\usepackage{amsthm}
\usepackage{color}
\usepackage{url}
\usepackage{listings}
\usepackage{float}  
\usepackage{booktabs}
\usepackage{caption}
\usepackage{subcaption}
\usepackage{xfrac}
\usepackage[colorlinks,bookmarksopen,bookmarksnumbered,citecolor=red,urlcolor=red]{hyperref}

\begin{document}

\title{Vectorized and performance-portable Quicksort}
\author{Mark Blacher \and Joachim Giesen \and Peter Sanders \and \href{mailto:janwas@google.com}{Jan Wassenberg}}

\maketitle
\abstract{Recent works showed that implementations of Quicksort using
vector CPU instructions can outperform the non-vectorized algorithms in
widespread use. However, these implementations are typically single-threaded,
implemented for a particular instruction set, and restricted to a small set of
key types. We lift these three restrictions: our proposed \emph{vqsort}
algorithm integrates into the state-of-the-art parallel sorter \emph{ips4o},
with a geometric mean speedup of 1.59. The same implementation works on seven
instruction sets (including SVE and RISC-V V) across four platforms. It also
supports floating-point and 16-128 bit integer keys. To the best of our
knowledge, this is the fastest sort for non-tuple keys on CPUs, up to 20 times
as fast as the sorting algorithms implemented in standard libraries. This paper
focuses on the practical engineering aspects enabling the speed and portability,
which we have not yet seen demonstrated for a Quicksort implementation.
Furthermore, we introduce compact and transpose-free sorting networks for
in-register sorting of small arrays, and a vector-friendly pivot sampling
strategy that is robust against adversarial input.}

\else

\documentclass[AMA,STIX1COL]{WileyNJD-v2}

\articletype{Article Type}%

\received{26 April 2016}
\revised{6 June 2016}
\accepted{6 June 2016}

\usepackage{microtype}
\usepackage{algorithm}
\usepackage{bigfoot}  
\usepackage{graphicx}
\usepackage{amsthm}
\usepackage{color}
\usepackage{url}
\usepackage{listings}
\usepackage{float}  
\usepackage{booktabs}
\usepackage{caption}
\usepackage{subcaption}
\usepackage{xfrac}
\lstset{language=C++,numbers=left,tabsize=2}

\raggedbottom

\begin{document}

\title{Vectorized and performance-portable Quicksort}

\author[1]{Jan Wassenberg}
\author[2]{Mark Blacher}
\author[2]{Joachim Giesen}
\author[3]{Peter Sanders}
\authormark{Wassenberg \textsc{et al}}

\address[1]{\orgname{Google Research}, \orgaddress{\country{Switzerland}}}

\address[2]{\orgdiv{Theoretical Computer Science}, \orgname{Friedrich Schiller University Jena}, \orgaddress{ \country{Germany}}}

\address[3]{\orgdiv{Institute of Theoretical Informatics, Algorithm Engineering}, \orgname{Karlsruhe Institute of Technology}, \orgaddress{\country{Germany}}}

\corres{Jan Wassenberg \email{janwas@google.com}}


\abstract[Summary]{
	Recent works showed that implementations of Quicksort using
vector CPU instructions can outperform the non-vectorized algorithms in
widespread use. However, these implementations are typically single-threaded,
implemented for a particular instruction set, and restricted to a small set of
key types. We lift these three restrictions: our proposed \emph{vqsort}
algorithm integrates into the state-of-the-art parallel sorter \emph{ips4o},
with a geometric mean speedup of 1.59. The same implementation works on seven
instruction sets (including SVE and RISC-V V) across four platforms. It also
supports floating-point and 16-128 bit integer keys. To the best of our
knowledge, this is the fastest sort for non-tuple keys on CPUs, up to 20 times
as fast as the sorting algorithms implemented in standard libraries. This paper
focuses on the practical engineering aspects enabling the speed and portability,
which we have not yet seen demonstrated for a Quicksort implementation.
Furthermore, we introduce compact and transpose-free sorting networks for
in-register sorting of small arrays, and a vector-friendly pivot sampling
strategy that is robust against adversarial input.
}

\jnlcitation{\cname{%
		\author{Wassenberg J}, \author{Blacher M}, \author{Giesen J},  and
		\author{Sanders P}.} (\cyear{2022}), \ctitle{Vectorized and performance-portable
		Quicksort}, \cjournal{Software: Practice and Experience}, \cvol{2017;00:1--6}.}

\maketitle

\fi  

\section{Introduction} \label{sec:intro}

Due to fundamental properties of current and expected future CPUs, including the
per-instruction energy cost, it is important for software to be designed to
utilize SIMD and/or vector extensions. Although SIMD and vector extensions
differ at the architecture level, since SIMD requires separate instruction encodings
for each vector size, we speak of SIMD/vector extensions interchangeably,
because software can be written in the same way for both. An instructive example
and the focus of this article is sorting, which is an important part of many
applications, including information retrieval. Replacing Quicksort from a
standard library by a vectorized Mergesort implementation can reduce energy
usage by a factor of six~\cite{sortEnergy}. Given these substantial gains in
energy and computational efficiency, it seems surprising that vectorized sorting
is not used much in practice. There are, however, some explanations for the so
far limited adoption of vectorized sorting.  Developing SIMD software involves
specialized domain expertise, including knowledge of the various instruction
sets, which is not a trivial requirement. For instance, Intel lists some 11,000
vector instructions and variants. This would be less of an issue if an
implementation could be written once and used widely. However, there are now at
least five major instruction sets across three architectures: x86 (AVX-512,
AVX2), Arm (NEON, SVE), and RISC-V (V extension). Thus, given the relative
scarcity of domain expertise, especially for the newer instructions sets, and
the nontrivial implementation complexity of state-of-the-art sorting algorithms,
it is not surprising that previous implementations of vectorized sorting are
specific to an instruction set. Furthermore, the availability of instruction
sets is problematic. For instance, AVX-512, proposed in 2013, is still not
widely supported. AMD CPUs may soon add support, whereas Intel's heterogeneous
Alder Lake platform disables AVX-512 for consistency between its two types of
cores. Thus, AVX-512 cannot be relied upon, except perhaps in some
supercomputers, where the hardware is typically known and rarely upgraded.
However, even the prior AVX2 instructions are only available in 86\% of a
survey's respondents' CPUs~\cite{steam}.

Here, we argue that it is no longer necessary to engineer software specific to
an instruction set. For linear algebra or `vertical' algorithms, where the
SIMD elements (lanes) are independent, autovectorization, that is, synthesizing
vector
instructions directly from C++ code by the compiler, is an appealing option.
However, re-ordering vector lanes, which is fundamental to sorting, is
infeasible via autovectorization~\cite{autovecPC}. In the absence of viable
compiler or language support, we use an abstraction layer, called Highway, over
platform-specific intrinsics (functions that map to vector instructions). In
C++, this is fairly straightforward as wrapper functions like, for instance,
\verb|Reverse| are easier to use than calling the corresponding
\verb|_mm512_permutexvar_epi16| intrinsic directly. Many such libraries have
been developed. However, some difficulties arise when choosing a set of
functions efficiently implementable on x86, Arm NEON, Arm SVE, and RISC-V V. The
latter two involve `scalable vectors' whose sizes are unknown at compile time,
which currently rules out some common C++ implementation techniques like
wrapping vectors in a class to enable member functions and specifying the vector
size as a compile-time constant. Furthermore, heterogeneous cloud servers and
client devices offer different instruction sets, requiring the application to
decide at runtime which instruction set is available for use. To the best of our
knowledge, our Highway C++ library is the only library that can handle `scalable
vectors' and check for the best available instruction set at runtime.
Application code is expressed by using calls to Highway functions, also known as
ops. This single implementation is automatically compiled for each requested
target by using the preprocessor to re-include the code, and \verb|#pragma|
statements to set the target architecture. Highway then chooses the appropriate
version at runtime. Here, we demonstrate the power of this approach by achieving
state-of-the-art performance results across multiple architectures from a single
implementation of a vectorized sorting algorithm.

Which sorting algorithms can be suitable for vectorization? Mergesort is
commonly used~\cite{merge2008,bertilMerge}, but typically requires $O(N)$ extra
storage. Mergesort also appears to be relatively slow in practice. An
implementation using wider AVX-512 vectors reports sorting
throughput~\cite{bertilMerge} only comparable to the speed of a vectorized
Quicksort using AVX2 vectors of half the size~\cite{blacherSEA}. Looking beyond
comparison-based sorts, Radixsort scatters keys to separate arrays. This has
generally, except for one algorithm tailored to the CRAY architecture and its
memory characteristics~\cite{vecRadix}, been implemented for single elements
without taking advantage of SIMD/vectors. Recent instruction sets (AVX-512, Arm
SVE, RISC-V V) include support for vectorized scatter. However, the case where
multiple lanes are to be scattered to the same array still remains a problem, perhaps to
be handled with x86-specific conflict detection instructions, or by providing
separate sets of arrays for each lane. The latter option results in numerous
arrays, which must either be grown as needed, or pre-allocated. To avoid
explicit checks whether arrays are full and to prevent committing large amounts
of memory, a previously described demand-allocation scheme using
virtual-memory~\cite{vmcSort} may be applicable. However, this is less portable
and practical. Taking these disadvantages of Mergesort and Radixsort into
account, we decided to design and implement a vectorized Quicksort, which is
also more cache-friendly and thus requires less memory bandwidth, an
increasingly scarce resource for shared-memory machines. Quicksort was already
vectorized for early supercomputers~\cite{qvecStar,qvecLevin} via compress
instructions. It took several decades until microprocessors were able to emulate
them using table-driven permutation instructions, which were applied to
Quicksort only several years ago~\cite{gueron}. Subsequent improvements include
in-place partitioning using AVX-512-specific compress
instructions~\cite{bramasAVX}, more robust pivot selection, and adding
sorting networks for small arrays~\cite{blacherSEA}. As mentioned before, all of
these works target a single instruction set. Meanwhile, the in-place sample sort
\emph{ips4o} constitutes the state-of-the-art for parallel, comparison-based
sorting according to a thorough experimental study~\cite{ips4o}, though its
current form does not take advantage of vector instructions.

This article introduces a vectorized implementation of Quicksort for 16--128 bit
elements, implemented using the Highway library's `portable', that is,
multiplatform-dependent, intrinsics. In addition to the core Quicksort
(recursive partitioning), our algorithm includes a sorting network ``base case''
and robust pivot sampling. The resulting \emph{vqsort} (vectorized
quicksort) is the fastest sorting implementation known to us for commercially
available shared-memory machines. Integrating \emph{vqsort} speeds up the
state-of-the-art \emph{ips4o} by a geometric mean of 1.59 and 2.89 in parallel
and single-core settings, respectively. We share the
production-ready open-source code~\cite{vqsortCode}.
Our specific contributions can be summarized as follows:
\begin{itemize}
	\item \textbf{generality:} support for 32/64-bit floating-point and
	16/32/64/128-bit integer keys in ascending or descending order,
	\item \textbf{performance portability:} the same implementation supports seven
	instruction sets with close-to native performance,
	\item \textbf{production-readiness:} \emph{vqsort} is
	open-sourced~\cite{vqsortCode}, tested, bounds-checked via compiler
	instrumentation, documented, and works with three major compilers.
\end{itemize}

\section{Vectorized quicksort} \label{sec:algo}

Conceptually, Quicksort is a simple algorithm. It recursively sorts arrays by
partitioning them with respect to some pivot element. The performance of
Quicksort crucially depends on the choice of the pivot element. In this section
we provide  a portable partitioning, faster than AVX-512-specific code
(Section~\ref{sec:algo:partition}), and a vectorized, cache-aware, robust pivot
sampling (Section~\ref{sec:algo:pivot}).

For small array sizes it pays off to switch to alternative sorting
algorithms/strategies such as sorting networks. Here, we provide a vectorized
sorting network which sorts 256 keys in several hundred CPU cycles
(Section~\ref{sec:algo:base}, and more details in Section~\ref{sec:network}).
Furthermore we support reverse sort order and 128-bit keys
(Section~\ref{sec:algo:base}).

A simplified C++ implementation of Quicksort is shown in
Algorithm~\ref{alg:recurse}, where \verb|T| refers to the key type,
\verb|[begin,end)| is the range to sort within the \verb|keys| array,
\verb|pivot| is a vector with all lanes set to the pivot obtained per
Section~\ref{sec:algo:pivot}, \verb|buf| is a preallocated buffer (sized
according to Section~\ref{sec:algo:base}), and \verb|rng| returns pseudorandom
unsigned 64-bit integers. The following sections explain the subroutines. For
details we refer the reader to the open-source code at
\url{https://github.com/google/highway/blob/master/hwy/contrib/sort/vqsort-inl.h}.
The set of Highway ops is documented at \url{https://github.com/google/highway/blob/master/g3doc/quick_reference.md}.

\begin{algorithm}
\caption{Quicksort recursion}
\label{alg:recurse}
\begin{verbatim}
void Recurse(T* keys, int begin, int end, V pivot, T* buf, R& rng) {
  int bound = Partition(keys, begin, end, pivot, buf);
  int num_left = bound - begin;
  int num_right = end - bound;

  if (num_right == 0) { // Degenerate partition
    V first, last;
    ScanMinMax(keys + begin, end - begin, buf, first, last);
    if (AllTrue(Eq(first, last))) return;
    return Recurse(keys, begin, end, first, buf, rng);
  }

  if (num_left <= NBaseCase()) {
    BaseCase(keys + begin, num_left, buf);
  } else {
    V next_pivot = ChoosePivot(keys, begin, bound, buf, rng);
    Recurse(keys, begin, bound, next_pivot, buf, rng);
  }
  if (num_right <= NBaseCase()) {
    BaseCase(keys + bound, num_right, buf);
  } else {
    V next_pivot = ChoosePivot(keys, bound, end, buf, rng);
    Recurse(keys, bound, end, next_pivot, buf, rng);
  }
}
\end{verbatim}
\end{algorithm}

After calling \verb|Partition| (Section~\ref{sec:algo:partition}), which returns
the starting index of the second partition, the remainder of
Algorithm~\ref{alg:recurse} is concerned with recursing to both partitions. It
is desirable to have the recursive function end with a call to itself. This
enables so-called `tail recursion', for which a jump instruction suffices,
avoiding the overhead of parameter passing and setting up a stack frame.

Note that we call \verb|ChoosePivot| (Section~\ref{sec:algo:pivot})
\textit{before} recursing, rather than inside \verb|Recurse|. This allows us to
choose a safe pivot whenever a degenerate (empty) partition is detected. With
some advance knowledge of the pivot and partitioning schemes (pivots are always
one of the input keys, and \verb|Partition| moves to the left any key equal to
the pivot), we are assured the left partition is never empty. Conversely, the
right partition is only empty if the pivot is equal to the last value in sort
order. If we again choose the same pivot, there is even a risk of infinite
recursion. Thus we must handle this case separately. The most likely cause is
that all keys in the current range are equal. This is quite common in
information retrieval applications, in which keys are often drawn from a small
subset of the possible values. We check for this by scanning through the keys
and computing their minimum and maximum value. This can be vectorized by
accumulating per-lane min and max, then `reducing' them to a single min/max
using Highway's \verb|Min/MaxOfLanes|. \verb|Eq| is a Highway op that returns a
mask indicating whether the inputs are equal, and \verb|AllTrue|
indicates whether all lanes of the mask are true. If the min and max are equal, then all
keys are also equal, and
thus already sorted, so we do not recurse further. Otherwise, the pivot was an
unlucky choice. Because we recursively use the median of three sampling, approximately
one third of the input must have been equal to the largest value. We only
observe this to happen with lower-entropy keys (e.g. uniform random 16-bit
integers within 32 or 64-bit elements). It happens less frequently with vectors
of size 16, which imply \verb|NBaseCase| $= 256$: large enough that a narrow
range of values within such a partition is unlikely. We therefore use a simple
heuristic that still guarantees forward progress: choosing the first key in sort
order as the pivot will partition off at least some keys, otherwise, they are
all equal, which was handled above.

Having hoisted \verb|ChoosePivot| out of the recursion, we also first check
whether the input is small enough to be handled directly in \verb|BaseCase|
(Section~\ref{sec:algo:base}). If so, the pivot would not be used anyway, and
guaranteeing a minimum input size is helpful for both \verb|Partition| and
\verb|ChoosePivot|.

\subsection{Partition} \label{sec:algo:partition}

Partitioning the input array is defined as moving elements which compare less
than or equal to the \verb|pivot| argument before the other elements. This
accounts for a large majority of compute time because it touches every key
during each of the expected $\log(n)$ passes over the input. \verb|Partition|
follows the basic approach of Bramas~\cite{bramasAVX}: an in-place bidirectional
scan using an AVX-512 instruction represented by the \verb|CompressStore|
Highway op. This accepts a vector and a mask as input and writes to contiguous
memory all vector lanes whose corresponding mask bits are true. To partition, we
simply \verb|CompressStore| elements at the left array side with the mask
obtained by comparing inputs to \verb|pivot|, and again on the right side with
the negated mask, advancing the write positions according to the number of
elements written, and stopping once they meet. Inputs are loaded from the left
or right side to maintain the invariant that all elements from the current loop
iteration could be stored either on the left or right sides. This entails
checking the `capacity' (difference between the read and write positions) on
either side, and loading from the one with less. To establish the invariant
before the loop, we begin by loading the first and last vectors of the input to
registers, to be partitioned after the loop.

In contrast to Bramas' explicit usage of AVX-512 instructions~\cite{bramasAVX},
the Highway op is portable. For AVX-512, \verb|CompressStore| maps directly to
an instruction except for 16-bit elements, which would require the not yet
widely available VBMI2 instruction set. On Arm SVE and RISC-V, the op stores the
result of a \verb|Compress| instruction to memory. For instruction sets without
per-lane masking, Highway emulates this operation by re-ordering the vector
according to a pattern loaded from a table, where the index is the concatenation
of the mask bits~\cite{gueron}.

We also find it is crucial for performance to unroll the partition
loop~\cite{blacherSEA}, possibly due to the conditional branch for deciding
whether to load the next elements from the left or right end of the array. We
also find branchless computations of the next address to be slower on a Skylake
CPU. Perhaps this is because the branch predictor sometimes guesses correctly,
thus reducing latency. Unrolling simply repeats each step in the loop, in our
case only four times as a compromise between code size, number of registers
required, and sufficient latency hiding. However, four vectors may exceed the
minimum guaranteed input size \verb|NBaseCase| (smaller inputs are handled by
\verb|BaseCase|). Thus we require an additional loop that partitions small
arrays. This is by definition not time-critical, so we adopt a simple approach:
overwriting the input via \verb|CompressStore| with the mask, and again
\verb|CompressStore| with negated mask to a buffer. Finally, we can append the
buffer contents to the current write position in the input. We must again handle
inputs which are not multiples of the vector size. As an extra complication, the
Highway op \verb|CompressStore| is allowed to overwrite memory after the valid
lanes. This simplifies the Arm SVE and RISC-V implementations by avoiding a
masked store. Such overwriting is fine for the padded buffer, but unacceptable
for the writes to the original input, for which we use the similar
\verb|CompressBlendedStore| op which avoids such overwriting, either with masked
stores, or by non-atomically `blending' the valid result with the previous
contents of memory following it. Note that we also use this small loop
(\verb|PartitionToMultipleOfUnroll| in the code) to handle remainders. This
simplifies the loop in \verb|Partition| by allowing it to assume its input is a
multiple of four vectors.

There is one further consideration: the first \verb|CompressBlendedStore| on the
right end of the array may read past its end. Because the address-sanitizer
feature of LLVM and GCC compilers checks whether vector loads are in-bounds,
this may trigger errors which terminate the program. To prevent this, we first
load the last vector of inputs into a register. After the remaining input has
been partitioned, we make space for one vector at \verb|writeL|, the first index
of keys in the right partition. This can be done with a single vector load and
store to the final vector by realizing that we wish to copy an entire vector
only if at least that many keys have been written in total to the right
partition.
Otherwise, we arrange for the right partition (less than a vector) to be stored
as the final elements of the last vector, by decreasing the load address by the
number of vector lanes less the right partition size. This is safe because we
ensure the input to \verb|Partition| consists of at least two vectors. With
space set aside, we are able to store the left keys of the final vector starting
at \verb|writeL|, and subsequently the right keys. The left keys overwrite the
space made above, i.e.\ duplicated keys. The next keys belong to the
right partition: either from the second vector of the right partition, or the
ones we just moved to the end. Thus \verb|writeL|, increased by the number of
left keys in the final vector, is the boundary between left and right
partitions, which \verb|Partition| returns.

The result of these efforts is portable code that outperforms AVX-512-specific
code~\cite{bramasAVX} by a factor of 1.7, more specifically, our 11644~MB/s vs.\
6891 as measured by their \verb|timePartitionAll<double>| benchmark for
$2^{24}$ items, compiled via \verb|clang++ -O2 -mavx512f|.

\subsection{Pivot selection} \label{sec:algo:pivot}

\verb|ChoosePivot| returns the pivot that will be passed to \verb|Recurse| and
thence to \verb|Partition|. Many published Quicksort implementations use medians
of constant-sized samples~\cite{adversary,badCases}. However, this traditional
approach would benefit from some adaptation for vectors and caches. Loading
elements from random array indices is possible using vector \verb|Gather| ops,
but these are expensive and emulated on the SSE4 and NEON instruction sets.
Furthermore, it seems wasteful to fetch an entire cache line from memory and
then only utilize one element. We instead load nine 64-byte chunks from random
64-byte-aligned offsets, and recursively reduce their elements to a single
median using medians of three as described below. The 64-byte chunk size
typically corresponds to the L1 cache line size.
Note that it would be onerous to detect the actual cache line size, and
unnecessary for correctness.

For each element index within a chunk, we determine the median of three elements
at same index within groups of three chunks we loaded. Medians of three can be
obtained with a sorting network consisting of four conditional swaps: (0,2)
(0,1) (1,2). The first grouping in this notation corresponds to replacing
elements 0 and 2 with their minimum and maximum, respectively. The latter two
groupings only require a total of two swaps because it suffices to correctly
order element 1, the median. We choose a sample size of three because sorting
network size is superlinear in the input size (e.g.\ already nine swaps for five
inputs).

When implemented using vector ops, this network is able to produce independent
results per lane, independently of the vector width. Thus we can iterate up to
the chunk size in units of the vector size, storing the resulting medians to a
buffer. Note that such a loop pattern is typical of vector-length-agnostic code,
which is preferred for the sake of portability.

Random bits are generated using a variant of SFC64~\cite{numpyIssues}, chosen
because it would support guaranteed-unique streams, though we did not use this
capability. To obtain offsets, we use a division-free modulo
algorithm~\cite{unbiasing} which only requires a single random draw per value.
This comes at the cost of some bias, which we expect to be acceptable. Some
numbers are generated less frequently than others, but the range of numbers,
i.e. chunk offsets, for sorting $2^{30}$ elements is $2^{24}$, implying a bias
of only $2^{-8}$.

We perform the above loads and median three times for a total of nine chunks
loaded and 192 bytes of medians. Given expected input sizes from $2^{20}$ to
$2^{30}$, this corresponds to roughly $\log(n)$ samples. We then reduce the
buffer to a single median, starting with the above approach to store the median
of three vectors from the input buffer to a second buffer. Once there are fewer
input elements than the vector size, we load single elements into vectors and
again compute medians with the same approach, but only store the first lane. The
remaining zero, one or two input elements are ignored. Finally, we swap the
buffers, recurse until fewer than three medians remain, and choose the first to
be the pivot.

Note that this constant-sized sampling strategy may lead to $O(N)$ recursions of
the main Quicksort~\cite{adversary} in the worst case. The C++ library
implementation in clang/LLVM was also vulnerable to this but has been
fixed~\cite{llvmIssues}. To prevent or rather detect such quadratic runtimes, we
impose a limit of $2 \cdot \log_2(n) + 4$ recursions. If exceeded, then we
switch to Heapsort, which we find to be ``only'' 20--40 times as expensive as
vectorized Quicksort (Table~\ref{tab:sort1M}). By contrast, binary Quicksort or
pivot switching~\cite{binaryQS,blacherSEA} may recurse up to 64~times for 64-bit
inputs, or even 1024~times ($\log_2$ of the maximum double-precision
floating-point exponent).

In practice, our sample is large enough to make the worst case extremely
unlikely to happen~\cite{badCases} except in adversarial settings, but detecting
and handling it only adds a few hundred bytes of code plus a well-predicted
branch, and guarantees $O(n \cdot \log(n))$ worst-case runtime.

If malicious input is possible and Heapsort would be unacceptably slow, a secure
random generator makes it infeasible for adversaries to predict the sampling
locations and thus cause a skewed pivot if the adversaries are not able to
access the random generator state in memory. Otherwise, adversaries could simply
clone and query the generator to determine the sample locations.

One such generator using hardware AES instructions as the round function of a
generalized Feistel network is indistinguishable from random unless the
adversary can perform more than $2^{64}$ work~\cite{randen}, which is more
expensive than the Heapsort fallback it might provoke. With thrifty use of this
more expensive generator (obtaining five 64-bit random values for the nine
32-bit chunk offsets), the sort is only 1--2\% slower if the
\verb|VQSORT_SECURE_RNG| option is enabled. This seems to be a reasonable cost
for the increase in robustness, but we disable it by default to avoid the
dependency on external code.

We also considered sampling a large fraction of the input, but this is
considerably slower on average. Finding the actual median deterministically
would also avoid any imbalance, but is reportedly an order of magnitude slower
than our \verb|Partition|~\cite{detSelect}, and thus cannot accelerate it.

\subsection{Base case} \label{sec:algo:base}

We now handle small arrays separately as a `base case' of the recursion, a
common optimization for Quicksort~\cite{blacherSEA,bingNetwork}. Sorting
networks built upon vector instructions can have much lower constant factors
than other algorithms because they execute fewer instructions and avoid
conditional branches. For moderate input sizes, this outweighs their higher $O(n
\cdot \log^2(n))$ complexity. With 256 or 512-bit vectors and 16--32~registers
commonly available, it is feasible to sort 64--256~elements within registers ---
an order of magnitude more than the five-element network found in LLVM's
Quicksort.

However, vector instructions entail handling input arrays that do
not evenly divide the vector size. Although instruction sets typically provide
some capability for only loading/storing valid lanes, AMD's x86 implementation
does not guarantee it can safely be used:\emph{``Exception and trap behavior for
	elements not selected for loading or storing from/to memory is implementation
	dependent. For instance, a given implementation may signal a data breakpoint or
	a page fault for doublewords that are zero-masked and not actually
	written''~\cite{amdRef}.} Thus the function \verb|BaseCase| begins by copying
the input range to \verb|buf| using the \verb|SafeCopyN| op, which either uses
masking or non-vector instructions to handle any remainder elements. To ensure
correct results, we then pad the buffer with neutral elements (the last value in
sort order) such that they remain in place while sorting. Our vectorized sorting
network (Section~\ref{sec:network}) can then load entire aligned vectors from
the buffer, and store the sorted results there. Finally, we again copy these
outputs to the original array.

Note that the buffer size is $O(1)$ with respect to the overall input to
Quicksort. Our sorting network reshapes $n$ inputs into a matrix of $r=16$ rows
and the smallest power of two $c \leq 16$ columns, such that $r\cdot c \geq n$
and $c$ elements fit within a vector. Thus \verb|NBaseCase| is $r\cdot c$ and
the buffer size must be at least $256$ elements, plus two vectors for padding in
case vectors are larger than $c$. We also reuse this buffer in
\verb|PartitionToMultipleOfUnroll| and \verb|ChoosePivot|. Thus it must also fit
at least nine vectors or four chunks plus two vectors. Because RISC-V (and to a
lesser extent Arm SVE) vectors may be large, the buffer size may exceed the
limit for stack allocation. Unfortunately, the C++ standard forbids
\verb|std::sort| from allocating memory dynamically. Thus \emph{vqsort} cannot
be used as a drop-in replacement on those platforms.

A framework for generating sorting networks from a domain-specific language has
been proposed~\cite{autoNet}. However, this relies on in-register transposition,
which is slower than the transpose-free networks that we propose and describe in
more detail in Section~\ref{sec:network}.

\subsection{Sort order and 128-bit keys} \label{sec:algo:order}

Note that user-specified comparators interact poorly with runtime dispatch
(choosing the sort implementation based on CPU capabilities). We
implement the latter by calling the best available implementation through an
indirect pointer. Unlike function templates such as \verb|std::sort|, this would
not allow us to inline user-specified functions. We expect that calling back to
a comparator through another function pointer would be expensive. If custom
comparisons are required, they can be inserted into a patched version of the
\emph{vqsort} source code, and exposed as a different sort function.

However, we do generalize comparisons to enable sorting in ascending or
descending order, which can be selected using a type-tag argument:
\verb|SortAscending| or \verb|SortDescending|. Our \emph{vqsort} implementation
is agnostic to the sort order because it builds upon an abstraction layer:
\verb|OrderAscending| and \verb|OrderDescending|. These define \verb|Compare|,
\verb|First|, \verb|FirstValue| for padding, and \verb|FirstOfLanes| (which is
equivalent
to the result of \verb|First| applied to successive lanes, but implemented using
Highway's \verb|MinOfLanes| reduction op). For every \verb|First*| there is also
a corresponding \verb|Last*|.

Recall that 128-bit keys (or 64-bit keys with 64-bit associated data) are
helpful for some information retrieval applications. SIMD/vector instruction
sets generally do not support 128-bit lanes natively. We can choose to split
them into 64-bit halves, such that one vector holds all lower halves of some
keys, or take advantage of the fact that Highway guarantees at least 128-bit
vectors to treat pairs of 64-bit lanes as unsigned 128-bit numbers. The former
is likely more efficient, but may require major changes to the memory layout of
applications. Thus we pursue the latter and find it to be about 0.7 times as
fast as native 64-bit sorts on x86, which is surprising given that 128-bit
comparisons require at least five instructions (taken care of by Highway). We
reproduce the x86 implementation in Algorithm~\ref{alg:lt128} for illustration;
the functions called there are all Highway ops.

\begin{algorithm}
	\caption{128-bit comparison using pairs of 64-bit lanes}
	\label{alg:lt128}
\begin{verbatim}
V eqHL = VecFromMask(d, Eq(a, b));
V ltHL = VecFromMask(d, Lt(a, b));
V ltLX = ShiftLeftLanes<1>(ltHL);
V vecHx = OrAnd(ltHL, eqHL, ltLX);
return InterleaveUpper(d, vecHx, vecHx);
\end{verbatim}
\end{algorithm}

\noindent This returns true in both lanes iff the upper lane is less, or the
upper lane is equal and the lower lane is less. The additional cost of pairs as
opposed to separate halves is due to the required interactions between the upper
and lower lanes. This is rather unusual and poorly supported in SIMD/vector
instruction sets, especially copying the upper lane to the lower lane in the
final line. However, it seems bearable, especially on x86. On AVX2, the
\verb|VecFromMask| op does not perform any work because masks are the same as
vectors. On AVX-512, this op does map to one instruction, but converting to a
vector enables fusing \verb|OrAnd| to a single ternlog instruction, and
surprisingly, shifting lanes has lower latency than shifting masks.

Because the emulated 128-bit comparisons also depend on the sort order, we
integrate them into the same abstraction, adding a layer to bridge the
differences between single-lane keys and pairs: \verb|KeyLane| vs.\
\verb|Key128|. These define functions such as \verb|Swap| (for \verb|HeapSort|),
\verb|SetKey| from a pointer, and others such as \verb|ReverseKeys| for use by
our sorting network. Shared code that depends on the order is grouped into
\verb|TraitsLane| and \verb|Traits128|. Finally, a \verb|SharedTraits| wrapper
class, abbreviated as \verb|st|, inherits from these and is passed to the
top-level \verb|Sort| function.

\section{Sorting networks} \label{sec:network}

For sorting arrays in the `base case' ($n \le 256$) we use sorting networks.
The building blocks of sorting networks are compare-and-exchange modules.
A compare-and-exchange module consists of two nodes. Each node receives a value. The two values in the module are sorted using a min and a max operation.
In a sorting network, the compare-and-exchange modules are combined in such a way that they always sort a fixed-length sequence of values.

Before we start sorting the arrays with sorting networks, we copy the elements
into an aligned buffer as described in Section~\ref{sec:algo:base}. We interpret
the buffer as a matrix in row-major order, where the number of columns
corresponds to the number of elements in a vector.  Our vectorization strategy
for sorting networks works as follows: First, the columns of the matrix are
sorted, then the sorted columns are directly merged with vectorized Bitonic
Merge networks~\cite{batcher1968}. We set the number of rows in the matrix to
16. Actually, any reasonable number of rows can be used, because, as we will
show, it is not necessary to transpose the matrix before merging the sorted
columns. We use a matrix with 16 rows because then Green's irregular sorting
network~\cite{knuth1998}, which has the smallest number of compare-and-exchange
modules for sorting 16 elements~\cite{codish2016}, can be applied directly to
sort the elements within the columns. Furthermore, 16 is a power of two. For
Bitonic Merge, a power of two as the number of elements to merge is particularly
efficient.

\begin{figure}[htbp] \centering \includegraphics[width=\textwidth]{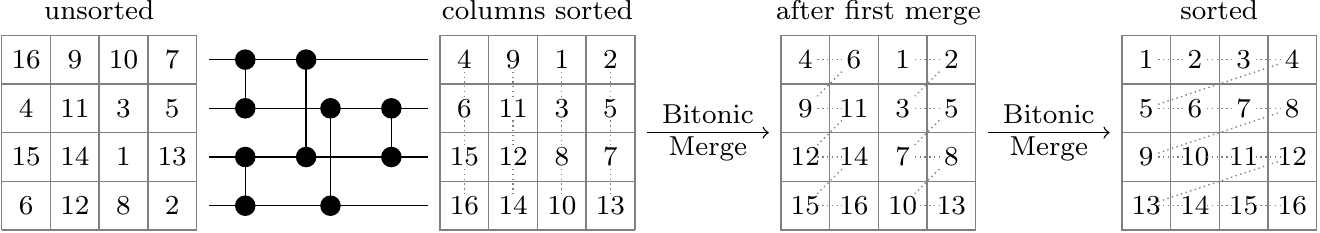}
	\caption{Sorting a $4\times4$~matrix. For sorting columns we use sorting
		networks with a minimum number of compare-and-exchange modules. To merge the
		sorted columns, we apply Bitonic Merge directly, without transposing the
		matrix.} \label{fig:sn:no:transposition} \end{figure}

To illustrate our vectorization approach to sorting networks, we discuss a
showcase example, where the capacity of elements in a vector is limited to four,
and a total of 16 elements needs to be sorted.  These restrictions result in a
$4\times4$~matrix. Figure~\ref{fig:sn:no:transposition}  provides a high-level
overview of sorting the $4\times4$~matrix and of our approach in general.

For sorting the values in each column, pairwise minima and maxima vector
operations are sufficient. Thus, sorting values within columns of a matrix with
sorting networks is particularly vector-friendly. Each vectorized
compare-and-exchange operation executes the same compare-and-exchange module in
all columns simultaneously. The number of instructions required to sort the
elements within the columns is therefore determined by the number of
compare-and-exchange modules in the sorting network. Sorting networks with a
minimum number of compare-and-exchange modules are particularly suitable for
sorting elements within columns. We use Odd-even Mergesort~\cite{batcher1968}
because its five compare-and-exchange operations for four elements correspond to
the lower bound~\cite{codish2016}.

Usually, after sorting the values column-wise the matrix is transposed, so that
the sorted column vectors become row vectors \cite{merge2008, bertilMerge}. We
avoid this transposition and start merging with vectorized Bitonic Merge
networks on the sorted columns themselves. First, the adjacent columns of the
matrix are merged. All two-column submatrices are sorted after the first merge
(see Figure~\ref{fig:sn:no:transposition}). Next, the adjacent two-column
submatrices are merged, resulting in sorted four-column submatrices. Since the
showcase has only four columns, the matrix is now sorted. If the matrix had
eight columns, another merge of the sorted four-column submatrices would be
required, and so on.

The basic idea behind merging sorted columns or sorted submatrices is to permute
the values of vectors so that the two nodes of each compare-and-exchange module
are placed under the same index in two different vectors. In other words: After
a permutation, the two nodes of a module are vertically aligned between two
different vectors. In our showcase, each vector contains four values, thus a
vectorized compare-and-exchange operation between two vectors executes at most
four  different modules. However, to demonstrate our merging strategy for sorted
columns, we use a vector size of two. A vector size of two is sufficient to
illustrate the first Bitonic Merge from Figure~\ref{fig:sn:no:transposition},
since the operations used are symmetric at larger vector sizes.

\begin{figure}[htbp]
	\centering
	\begin{subfigure}{\textwidth}
       \centering
		\includegraphics[width=\textwidth]{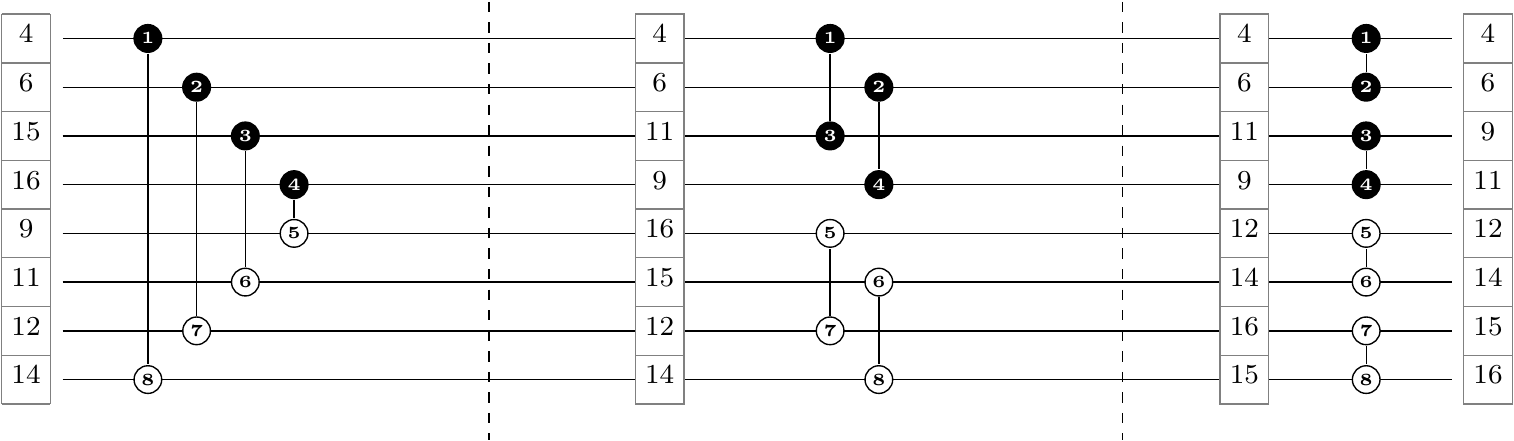}
		\caption{scalar Bitonic Merge}
		\label{fig:sn:merge:scalar}
	\end{subfigure}
    \hfill
    \vspace{2mm}
	\begin{subfigure}{\textwidth}
        \centering
		\includegraphics[width=\textwidth]{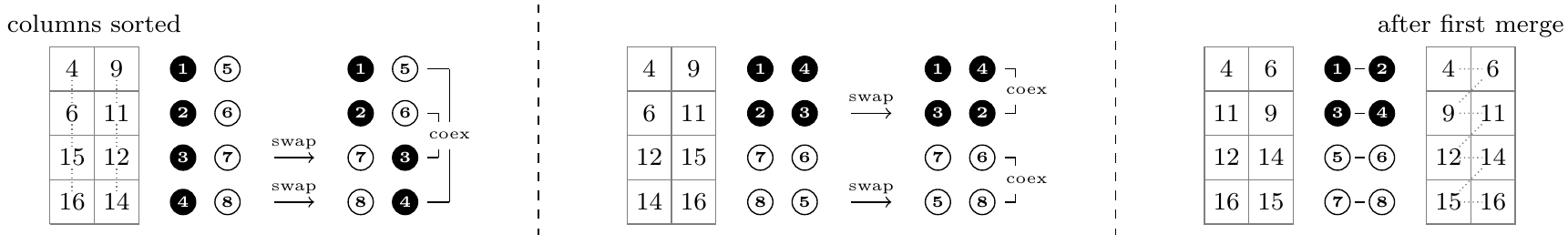}
		\caption{vectorized Bitonic Merge}
		\label{fig:sn:merge:vectorized}
	\end{subfigure}
	\caption{Bitonic Merge of two sorted arrays. Each array contains four values. }
	\label{fig:sn:merge}
\end{figure}

Figure~\ref{fig:sn:merge:scalar} contains the scalar version for merging two
sorted four-element subarrays. In the first merge step the compare-and-exchange
modules (1,8), (2,7), (3,6)  and (4,5) are executed. To execute  the same
modules using vectorized compare-and-exchange operations (coex) in a
$4\times2$~matrix, we first swap the adjacent elements of the last two vectors
(see Figure~\ref{fig:sn:merge:vectorized} left). In the second merge step the
compare-and-exchange modules (1,3), (2,4), (5,7)  and (6,8) are executed. In the
vectorized version (Figure~\ref{fig:sn:merge:vectorized} center), we swap the
adjacent elements of the  second and fourth vectors before executing  the  two
vectorized compare-and-exchange operations. In the last step, adjacent elements
must be compared and exchanged. If we had more than two values per vector, the
last merge step of Figure~\ref{fig:sn:merge:vectorized} (right) could also be
vectorized. But, then only half of the capacity of the vectors would be used,
since the nodes of a module are within one vector and not distributed over
different vectors. For our example with two elements per vector, scalar
compare-and-exchange operations can be used in the last merge step instead.

Figure~\ref{fig:sn:merge:vectorized} shows how to apply Bitonic Merge to sorted
columns. Similarly, Bitonic Merge can be applied directly to merge sorted
submatrices. The basic idea remains the same: the values of the vectors are
permuted so that the two nodes of each module are under the same index in two
different vectors, and then vectorized compare-and-exchange operations can be
performed between vectors representing the same modules but opposite nodes.

\section{Performance evaluations} \label{sec:perf}

\subsection{Memory bandwidth is the bottleneck}

\begin{figure}[h!]
	\centering
	\includegraphics[width=0.65\textwidth]{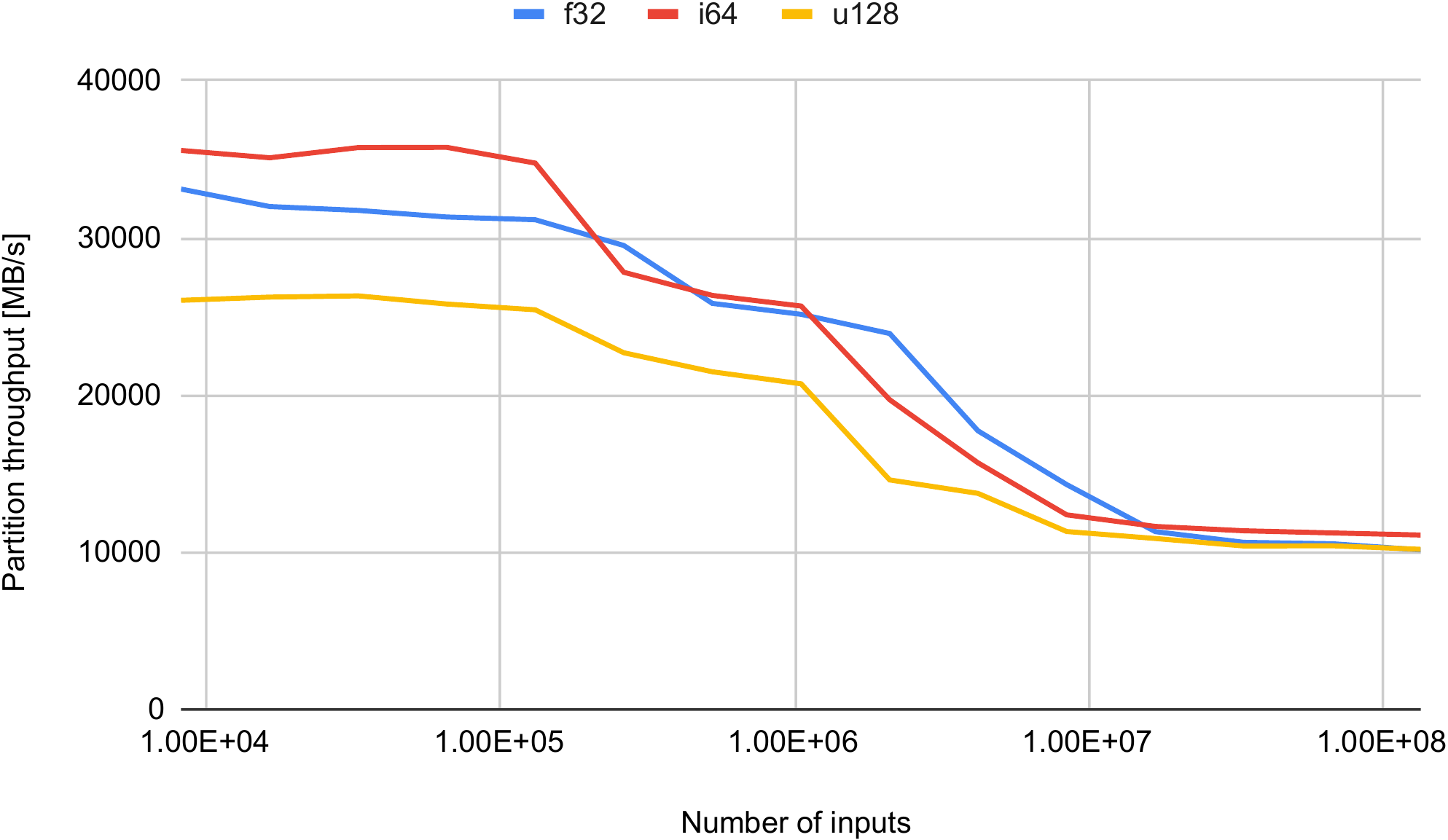}
	\caption{Partition throughput [MB/s] for various data types on a single 3~GHz
		Xeon Gold 6154 core, by number of inputs.}
	\label{fig:part1}
\end{figure}

So far, we have focused on efficient use of vector instructions. However, in our
experience, memory bandwidth is usually the limiting factor for the performance
of vectorized software. Here also, we find that partitioning cache-resident data
is two to three times as fast as partitioning large amounts of data in memory
(Figure~\ref{fig:part1}). This holds even for 128-bit keys, which as we saw in
Algorithm~\ref{alg:lt128} require at least five instructions per comparison.

\begin{figure}[h!]
	\centering
	\includegraphics[width=0.65\textwidth]{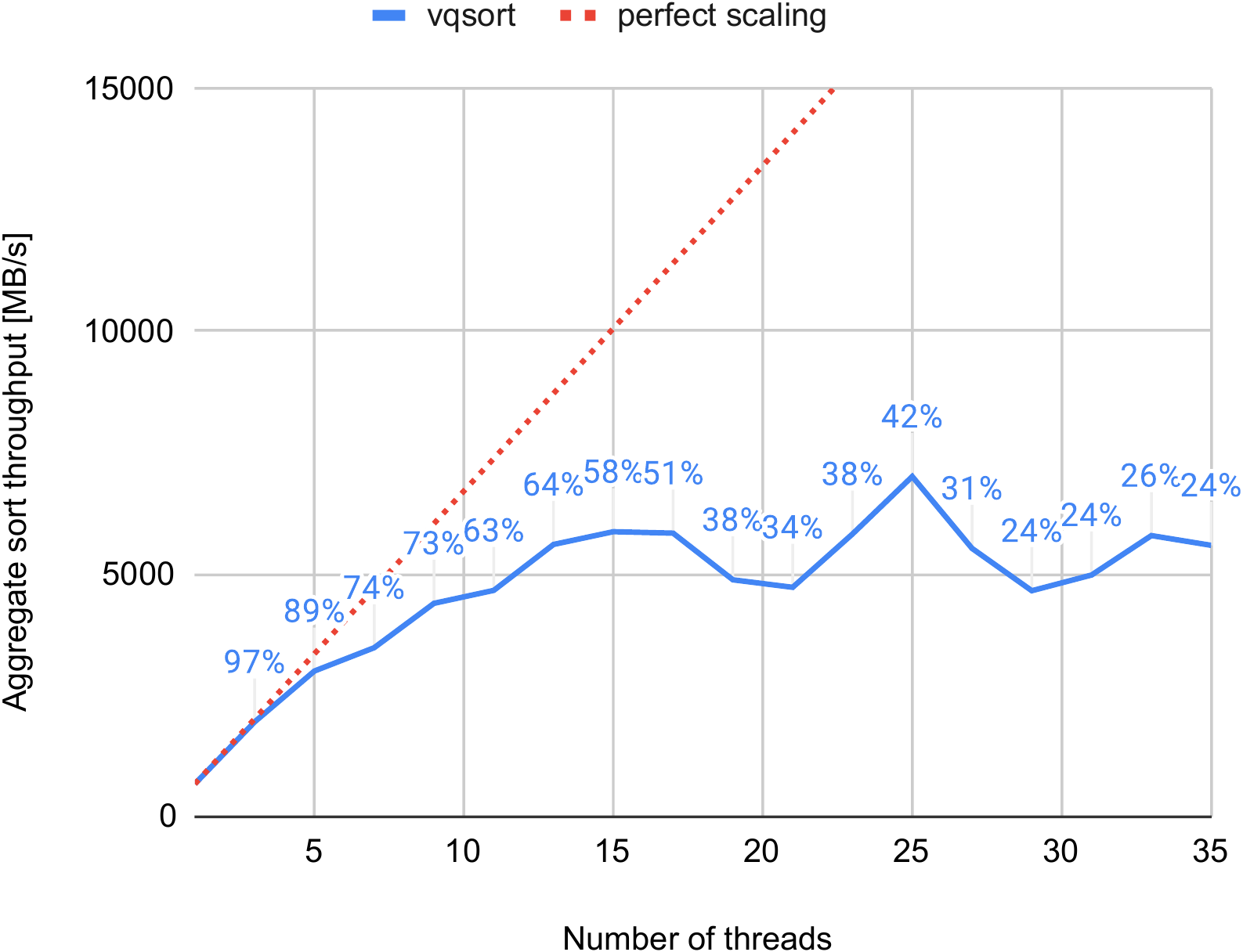}
	\caption{Aggregate sort throughput [MB/s] for 100M uniform random i64 keys on
		two Xeon Gold 6154 CPUs at 3~GHz, by number of independent instances, plus
		parallel efficiency.}
	\label{fig:par:vq}
\end{figure}

\noindent From this, we can conclude that Skylake CPUs are surprisingly
efficient at executing vector instructions, but their single-core memory
bandwidth seems under-provisioned relative to the vector capabilities. This gap
widens when considering multiple cores. The aggregate sort throughput of
concurrent sorts of 100M items (far exceeding the L3 cache size), running on a
simple thread pool, plateaus after using only 40\% of the cores
(Figure~\ref{fig:par:vq}).

\noindent We speculate that the memory bandwidth provisioning is a choice rather
than an unavoidable constraint. Non-vector workloads are less likely to expose
this limitation. Assuming SPEC 2017 benchmarks are representative, their 1.5
instructions per cycle and 0.4 loads per retired \textmu op~\cite{specChar}
impute bandwidth requirements of 7.2 or 14.4 GB/s (for 32- or 64-bit loads) at
3~GHz, for which Skylake cores seem adequately provisioned. However, as we have
seen, vector workloads such as \verb|Partition| (which includes non-negligible
computation per load) can utilize more than twice that per core. Fujitsu's A64FX
demonstrates that it is feasible to integrate High-Bandwidth Memory (HBM) into
CPUs, enabling about 1~TB/s bandwidth per chip, shared among 48~cores. The
upcoming Intel Sapphire Rapids CPU with HBM may also deliver similar bandwidth
increases. However, these are supercomputer or server-class CPUs and not yet
widely used or available. As an alternative or complement to hardware
improvements, we also consider algorithm-level changes.

\subsection{More bandwidth-friendly algorithms}

Recall that Quicksort only splits $N$ inputs into two partitions, requiring
about $\log_2(N)$ recursions. If we instead scatter inputs into $K$ partitions,
the base of the logarithm changes to $K$, which seems promising. However, our
method of compressing vector lanes and storing to each partition seems
unsuitable for $K \ge 8$ and current vector lengths of 512-bits. Given 64-bit
keys, we would only be writing one key on average to each partition, and
\verb|CompressStore| can only execute every other clock cycle on Skylake. Thus
the throughput would be limited to 64 bytes per 32~cycles, or 6~GB/s at 3~GHz,
which is far below the Skylake L3 cache bandwidth as seen in
Figure~\ref{fig:part1}. That leaves $K=4$, which was previously found to be
helpful~\cite{multiPiv} in a non-vectorized context. That algorithm can benefit
from conditional branches, which allow some comparisons to be skipped. However,
we prototyped vectorized compress with $K=4$ and found it to reach about half
the speed of $K=2$, thus negating the gain from halving the number of
recursions.

\begin{figure}[h!]
	\centering
	\begin{subfigure}[b]{0.49\textwidth}
		\centering
		\includegraphics[width=\textwidth]{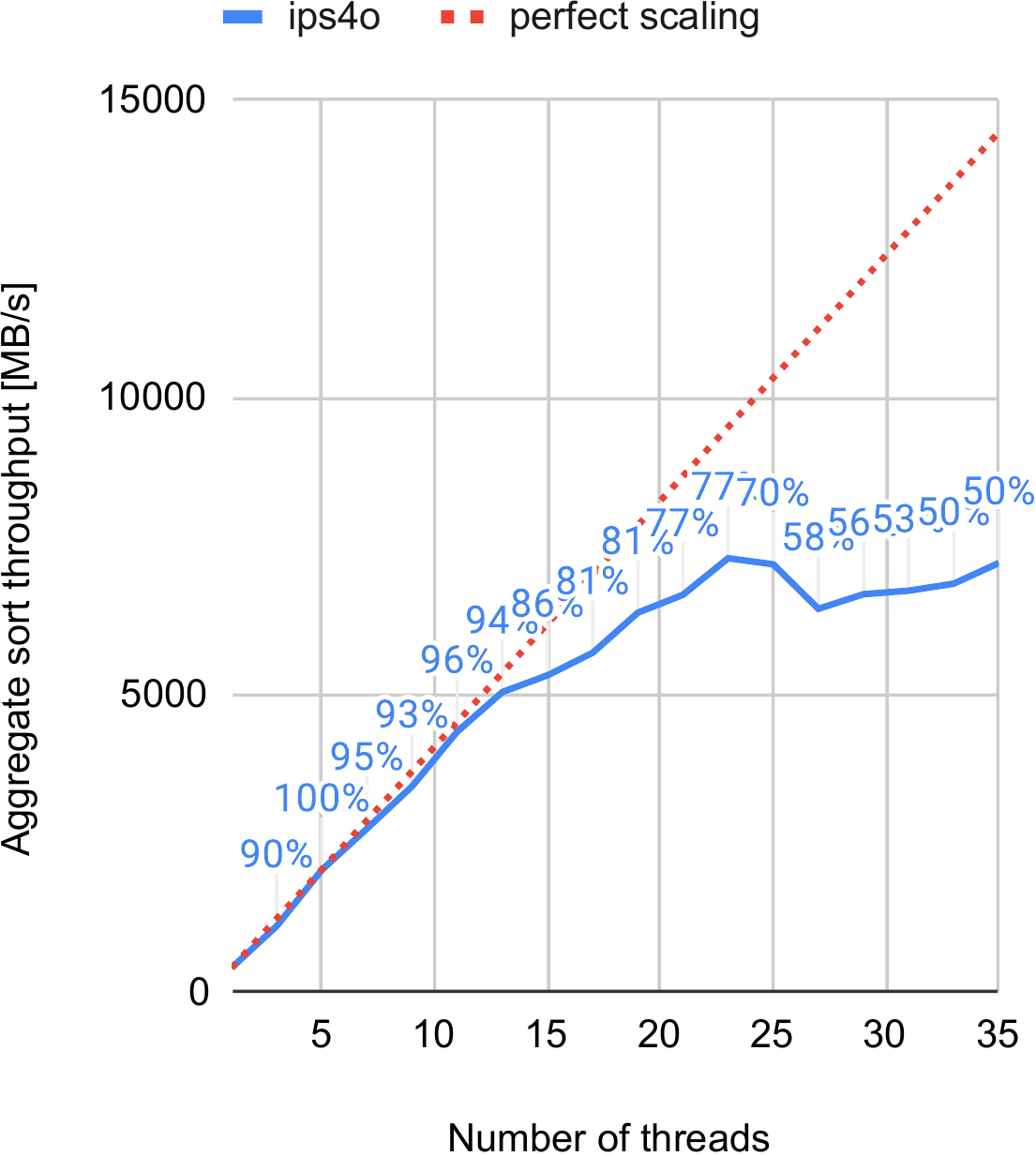}
		\caption{ips4o}
		\label{fig:par:ip}
	\end{subfigure}
	\hfill
	\begin{subfigure}[b]{0.49\textwidth}
		\centering
		\includegraphics[width=\textwidth]{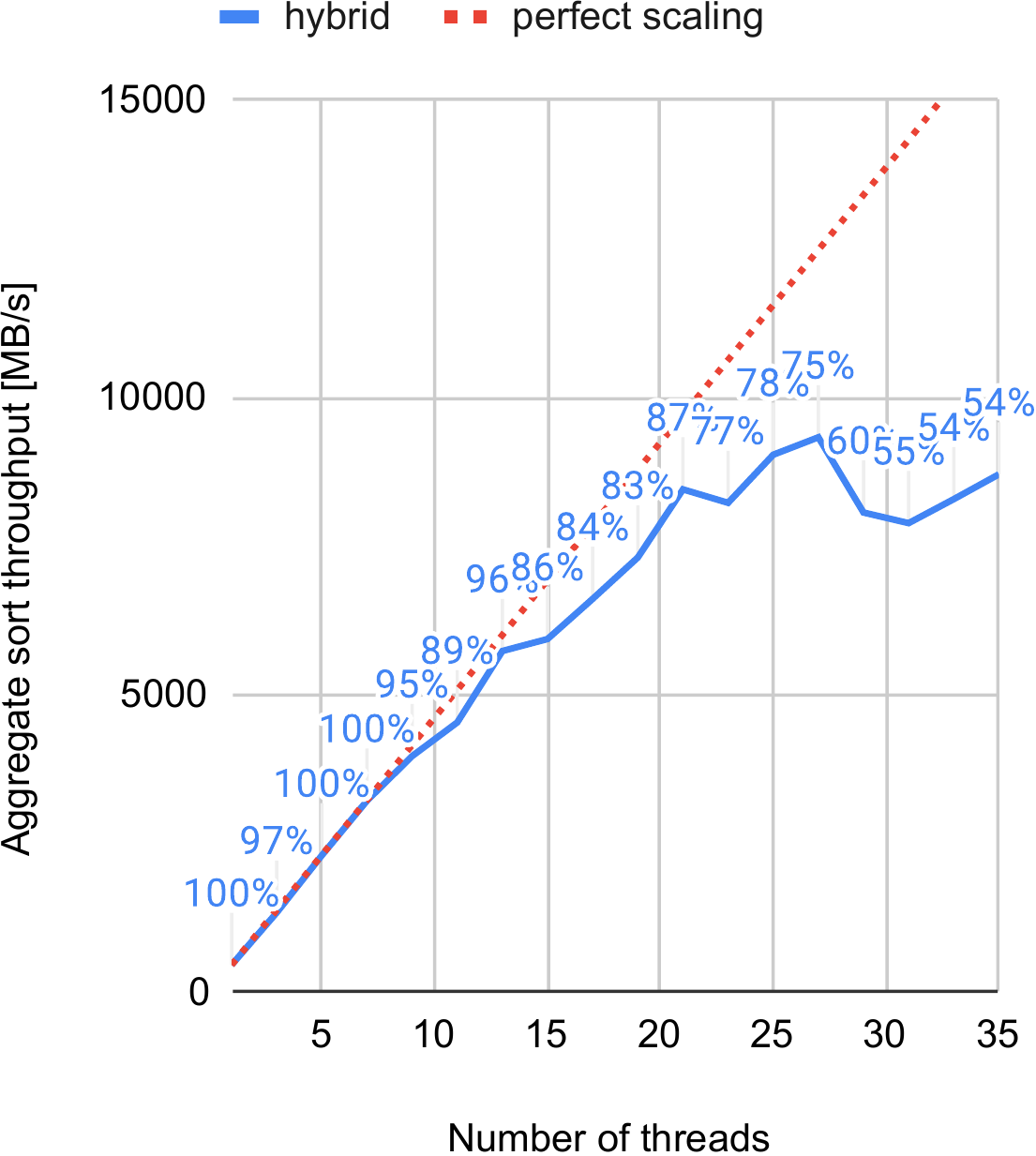}
		\caption{hybrid of \emph{vqsort} and \emph{ips4o}}
		\label{fig:par:hy}
	\end{subfigure}
	\caption{Aggregate sort throughput [MB/s] for 100M uniform random i64 keys on
	two Xeon Gold 6154 CPUs at 3~GHz, by number of independent instances, plus
	parallel efficiency.} \label{fig:par}
\end{figure}

If vectorized multi-pivot is unhelpful, what about the extreme case of
Samplesort, which is essentially a very large ($K=256$) generalization? As
expected, \emph{ips4o} scales better than \emph{vqsort}
(Figure~\ref{fig:par:ip}) because it requires fewer passes over the data, thus
reducing pressure on the shared memory system. However, in absolute terms, it is
slower in aggregate for less than 19~threads. For a possible explanation, we
note that \emph{ips4o} executes nearly five times as many instructions (5.2T
versus 1.1T) because its current form does not take advantage of vector
instructions.  As mentioned in the introduction, it may be possible to
accelerate \emph{ips4o} using vector instructions for scattering keys, or also
slightly accelerating the comparisons used to classify keys into buckets.
However, this appears to be difficult given our goal of a portable and
vector-length-independent algorithm. We instead pursue a simpler approach. Given
that \emph{ips4o} scales better, but has lower single-core throughput, we can
switch to \emph{vqsort} after several initial recursions of \emph{ips4o}. To
this end, we simply change \verb|baseCaseSort| to call \emph{vqsort}, set
\verb|IPS4OML_BASE_CASE_SIZE| to 8192, and tweak \verb|IPS4OML_BLOCK_SIZE| and
\verb|IPS4OML_UNROLL_CLASSIFIER| to 1024 and 6, respectively. This improves
scalability (Figure~\ref{fig:par:hy}) and the geometric mean of speedups
relative to \emph{ips4o} is 1.18. One may consider that underwhelming. Although
this benchmark seems representative of applications that divide their tasks into
independent shards, the memory bandwidth bottleneck limits the speedup that can
be observed. Measurements from the CPU uncore at one second granularity confirm
that each socket sees up to 74~GB/s read+write traffic, about 80\% of the value
observed when running the multithreaded STREAM benchmark v5.10. Thus we also
measure in two other settings. First, a single instance of \emph{ips4o}'s
parallel mode using 16 threads is much less bandwidth-intensive, about 12 GB/s
according to the same uncore measurements. The geometric mean of the speedups of
our hybrid vs.\ \emph{ips4o} is 1.59 (Table~\ref{tab:parsort100M}). Second,
although a single core with
near-exclusive usage of the L3 cache is likely not representative of server
workloads, we include the results for completeness. Our hybrid is \textbf{2.89}
times as fast as \emph{ips4o} (geometric mean), though still only 39-70\% the
speed of \emph{vqsort} (Table~\ref{tab:sort1M}). \emph{vqsort} using AVX-512 is
in turn \textbf{18.9}, \textbf{20.0}, \textbf{9.6}, and \textbf{8.9} times as
fast as the LLVM
C++ library implementation of \verb|std::sort|. \emph{vqsort} using AVX-512 is
1.5 to 2.0 times as fast as on AVX2, which has half the vector width but may
permit slightly higher CPU clock frequencies.

\begin{table}[h!]
	\caption{Aggregate sort throughput [MB/s] using 16 threads on one Xeon Gold 6154 CPU for various algorithms and data types.100M uniform random keys, turbo disabled}
	\centering
	\begin{tabular}{r r r}
	  \toprule
	    & \emph{ips4o} & \textbf{hybrid} \\
	   \midrule
	   f32 & 2385 & \textbf{2552} \\
	   i32 & 1512 & \textbf{2904} \\
	   i64 & 2710 & \textbf{4779} \\
	  u128 & 2703 & \textbf{4719} \\
	  \bottomrule
	\end{tabular}
	\label{tab:parsort100M}
\end{table}

\begin{table}[h!]
	\caption{Single-core sort throughput [MB/s] on one Xeon Gold 6154 CPU for various algorithms and data types. 1M uniform random keys; VQ256 denotes vqsort using AVX2}
    \centering
    \begin{tabular}{r r r r r r r}
      \toprule
         & \emph{ips4o} & hybrid & \textbf{vqsort} & VQ256 & std & Heapsort \\
      \midrule
       f32 & 142 & 445 & \textbf{1135} & 715 &  60 & 29\\
       i32 & 152 & 495 & \textbf{1161} & 795 &  58 & 28\\
       i64 & 284 & 700 & \textbf{1137} & 628 & 118 & 53\\
      u128 & 287 & 791 & \textbf{1142} & 569 & 128 & 53\\
      \bottomrule
    \end{tabular}
    \label{tab:sort1M}
\end{table}

\subsection{Performance portability}
Performance portability entails not only running on other platforms, but also
reaching a high degree of efficiency. We show that this is achievable
on recent x86 CPUs as well as on other platforms with weaker vector units.
We measured the same source
code and benchmark on an Apple M1~Max system (Table~\ref{tab:m1}). Note that the
results are not directly comparable with the Xeon because the M1's clock rate
differs (3.2~GHz). Even with the M1's 128-bit vectors and the
older NEON instruction set, we observe a 3-8x speedup over the standard library.
Thus \emph{vqsort} appears to be practical and useful on multiple architectures
and instruction sets.

\begin{table}[h!]
	\caption{Single-threaded sort throughput [MB/s] on M1~Max for 1M uniform random keys.}
	\label{tab:m1}
	\centering
	\begin{tabular}{r r r r}
		\toprule
		& \textbf{vqsort} & std & Heapsort \\
		\midrule
		 f32 & \textbf{498} &  63 & 10\\
		 i32 & \textbf{499} &  76 & 13\\
		 i64 & \textbf{471} & 151 & 69\\
		u128 & \textbf{466} & 151 & 69\\
		\bottomrule
	\end{tabular}
\end{table}

\section{Limitations} \label{sec:limitations}

To facilitate vectorization, we imposed a constraint on sort keys, namely,
that they be 16/32/64-bit integers, floating-point numbers, or pairs of 64-bit
numbers representing a 128-bit integer. (Note that x86
CPUs prior to Icelake are not able to efficiently re-order 8-bit elements across
a register, which requires the VBMI instruction set.) This constraint excludes
some applications that need to sort tuples or large items with custom
comparators. However, we argue that sorting numbers is still in widespread use:
`columnar' databases store the values of each column contiguously, and fields
are often encoded as numbers. Database query engines typically require a stable
sort (order-preserving among equivalent keys), which Quicksort is not, but can
be made so by appending a unique (row) identifier to the least-significant bits
of the key. Thus we have one important target application, typically involving
64-bit numbers.

To understand the impact, we surveyed uses of sorting in Google's production
workloads and found that sorting numbers is actually more costly in total than
sorting strings or user-defined
types including tuples. Our methodology starts by searching in Google's entire
source code depot for occurrences of \verb|std::sort| and the wrapper function
\verb|absl::c_sort|. A small fraction of these are excluded based on their
filename
(e.g.\ non-source files) or path (e.g.\ compiler test suites). We then exclude
the vast majority whose directories do not account for a relevant number of
samples in Google-wide CPU usage measurements. This leaves several hundred
occurrences, which are still too numerous for manual inspection. We further
filter out calls (about half) which have an extra comparator argument. Note that
some of them may define a lexicographical ordering within 128 or fewer bits of
data, which could be supported by \emph{vqsort}. However, this would be
laborious to prove, so we exclude them from our analysis. We then manually
inspect the code, finding that the total CPU time for sort calls with up to
128-bit keys outnumbers the total for other sorts (e.g.\ strings and tuples) by
a factor of two. Although we are surprised by this result, the straightforward
and mostly automated methodology makes us reasonably confident that the analysis
is valid. However, there
is one major caveat: we only find calls to the standard library sort. Other
potential sort-like algorithms such as tournament sort are not included in the
analysis.

We remark that vectorizing sorts with custom comparators is still possible.
\verb|Partition| already calls a comparison function. The larger change required
would be to replace \verb|Min/Max| in \verb|Partition| with comparisons and
conditional swaps, which we leave for future work.

\section{Conclusions} \label{sec:conclusion}

We used the Highway cross-platform abstraction layer for implementing
\emph{vqsort} (vectorized Quicksort) and utilizing the most efficient
instructions
available on the current CPU. The algorithm features a new recursive sorting
network for up to 256 elements that mitigates the previously
reported~\cite{bingNetwork} problem of excessive code size, and also a
vector-friendly pivot sampling, with the surprising result that robustness
versus adversarial input increases CPU cost by less than 2\%.

Our measurements indicate that \emph{vqsort} makes very good use of AVX-512
instructions. To the best of our knowledge, it is the fastest sort for
individual (non-tuple) keys on AVX2 and AVX-512, outperforming the standard
library's sort by factors of 10 to 20. When using Arm NEON on Apple M1 hardware,
we observe a 3-8x speedup versus the standard library. We are currently only
able to test the
code for SVE and RISC-V via emulators which do not predict performance.
SVE benchmarking may be feasible once Amazon's Graviton3 CPUs are publicly
accessible. However,
we can reasonably expect good results because both provide native compress
instructions. Integrating \emph{vsort} into the state-of-the art parallel sorter
\emph{ips4o}~\cite{ips4o} yields a speedup of 1.59 (geometric mean).

In contrast to previous works, which can be seen as proofs of concept, we have
focused on \textbf{practical usability} (support for 32/64-bit floating-point
and 16/32/64/128-bit integer keys in ascending or descending order), and
\textbf{performance portability} (supporting seven instruction sets with
close-to native performance).

\pagebreak
\bibliography{references}%

\end{document}